\documentclass[acus]{JAC2000}

\usepackage{graphicx}

\begin{document}
\title{Estimates of Dispersive Effects in a Bent NLC Main Linac\thanks{Work
supported by the U.S. Department of Energy}}

\author{M. Syphers and L. Michelotti, Fermilab, Batavia, IL 60510, USA}

\newcommand{\bs}{\char '134 }     
\newcommand{\fig}[1]{Figure~\ref{fig:#1}}

\newcommand{\secref}[1]{Section~\ref{sec:#1}}

\newcommand{\eq}[1]{Eq.(\ref{eq:#1})}

\newcommand{\eqs}[1]{Eqs.(\ref{eq:#1})}

\newcommand{\equ}[1]{Equation~(\ref{eq:#1})}

\newcommand{\textmath}[1]{\mbox{$#1$}}

\newcommand{\undl}[1]{\underline{#1}}

\newcommand{\dblundl}[1]{\underline{\underline{#1}}}

\newcommand{\cpp}{C\raisebox{.2ex}{++}\ }

\newcommand{\extraspace}{

  \vspace{0.5\baselineskip}  

  \newline

  }

\newcommand{\doublespace}{

  \vspace{\baselineskip}  

  \newline

  }

\newcommand{\triplespace}{

  \vspace{2\baselineskip}  

  \newline

  }

\newcommand{\binomcoeff}[2]{
  \left(
  \begin{array}{c}
  #1 \\
  #2
  \end{array}
  \right)
  }
\newcommand{\dotprod}[2]{
  \underline{#1} \cdot \underline{#2}
  }
\newcommand{\since}{
  \raisebox{.8ex}{.}.\raisebox{.8ex}{.}
  }
\newcommand{\therefore}{
  .\raisebox{.8ex}{.}.
  }
\newcommand{\theset}[2]{
  \{ \, #1 ~|~ #2 \, \}
  }
\newcommand{\tsfrac}[2]{
  \frac{\textstyle #1}{\textstyle #2}
  }
\newcommand{\ts}{\textstyle}
\newcommand{\norm}[1]{
  \parallel \, #1 \, \parallel
  }
\newcommand{\smbfmtrx}[1]{
  \dblundl{#1}
  }
\newcommand{\mtrx}[2]{
  \left(
    \begin{array}{#1}
    #2
    \end{array}
  \right)
  }
\newcommand{\commutator}[2]{
  [ \, #1 , \, #2 \, ]
  }
\newcommand{\poissonbracket}[2]{
  \{ \, #1 , \, #2 \, \}
  }
\newcommand{\dragt}[1]{
  { : \! #1 \! : }
  }
\newcommand{\xd}{{\bf d}}

\maketitle

\begin{abstract} 
An alternative being considered for the Next Linear Collider (NLC) is not to
tunnel in a straight line but to bend the Main Linac into an arc so as to
follow an equipotential. We begin here an examination of the effects that
this would have on vertical dispersion, with its attendant consequences on
synchrotron radiation and emittance growth by looking at two scenarios: a
gentle continuous bending of the beam to follow an equipotential surface,
and an introduction of sharp bends at a few sites in the linac so as to
reduce the maximum sagitta produced.
\end{abstract}

\section{Continual gentle bends}
%
%
In our first scenario,
the Main Linac remains 
as close as possible
to an equipotential surface.
Minimalism suggests that 
we try bending the beam by vertically
translating already existing NLC quadrupoles,
without introducing new elements or additional magnetic fields.
We thus propose that steering be accomplished 
by precisely aligning all the quads
``level'' along the equipotential 
and then raising the vertically defocusing (D) quadrupoles
to steer the beam through the centers of the
vertically focusing (F) quads.
\footnote{
The usual convention is for ``F'' (``D'') 
to indicate a horizontally focusing (defocusing ) quadrupole.
We do the opposite here, because we are considering
dynamics only in the vertical plane.
}
Bending at the D quad locations will
minimize the generated dispersion.

To estimate the order of magnitude 
of dispersion produced by such an arrangement,
we calculate (a) assuming a periodic sequence of magnets
while (b) neglecting the effects of acceleration \cite{mich:cr2st}
and (c) keeping only leading terms in the bend angle.  Our results will be
reasonably correct provided that upstream injection into the Main Linac
is redesigned to match the new arrangement.  Further details of the
calculation and others discussed in this paper are documented elsewhere.
\cite{mich:FNAL690}

\begin{figure}[ht]
\centering
\includegraphics*[width=90mm]{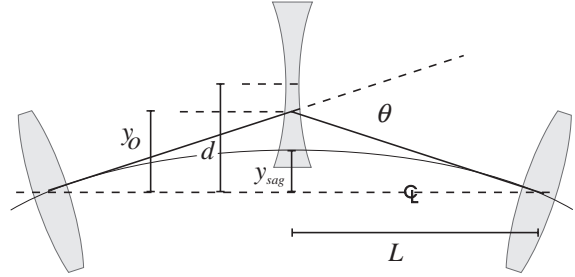}
\caption{
\label{fig:r79g}
Description of parameters for describing the 
CR thin quad calculations.
        }
\end{figure}

\fig{r79g} shows the physical layout of quadrupoles
and identifies the geometric parameters.
It is practical to write the vertical offset of the quadrupole relative
to the equipotential, $d-y_{\rm sag}$. In terms of the distance between
quadrupoles, $L$, the local betatron phase advance per cell, $\mu$, and the
radius of the earth, $R$, this offset is
\begin{equation}
d - y_{\rm sag} 
  =  \frac{L^2}{R} \left( \, \frac{1}{2} + \frac{1}{\sin(\mu/2 )} \,
\right)
\label{eq:ih49bn}
\end{equation}
To make a numerical estimate of this offset
at the high energy end of the linac,
we take 
\textmath{L \approx 19 \, \mathrm{m},}
\textmath{\mu \approx \pi / 2,}
and
\textmath{R \approx 6400 \, \mathrm{km}.}
\eq{ih49bn} then yields
\textmath{d - y_{\rm sag} \approx 108 \, \mu \mathrm{m}.}
%
%
The dispersion can be estimated easily 
using two observations:
(1) in passing through a thin bending magnet,
the slope of the dispersion function, $D'$, changes
by an amount approximately equal to the bend angle;
(2) by symmetry, the dispersion attains its maximum (minimum) 
value at the center of the focussing (defocussing) quadrupole.
The values of the dispersion function at the thin-lens quadrupoles will be
\begin{eqnarray*}
D_{\rm min} & = & \frac{L^2}{R \sin^2(\mu/2)} \left( \, 1 - \sin(\mu/2)  \,
\right) \; \; ,
\\
D_{\rm max} & = & \frac{L^2}{R \sin^2(\mu/2)} .
\end{eqnarray*}
%
%
%
%

Using the same parameters as before, 
this provides the numerical estimate,
at the high energy end of the linac,
\begin{displaymath}
D_{\rm min} = 0.032 \, \mathrm{mm,}\ \ \ D_{\rm max} = 0.11 \, \mathrm{mm.}
\end{displaymath}
\noindent
If we take a large \textmath{\Delta p / p \approx \Delta E / E = 0.02,}
because of BNS damping, 
and assume that the ``invariant emittance''
\textmath{\gamma \epsilon_y / \pi \approx 100\, \mathrm{nm}}
and \mbox{$\beta_y \approx 40$ m}
at a point where the electron's energy is
\mbox{$E = 100$ GeV,} then 
$D_{\rm max} \cdot \frac{\Delta p}{p} = 2.2 \, \mu \mathrm{m}$
 compared to
$\sigma_y = \sqrt{\beta_y \epsilon_y / \pi} = 4.6 \, \mu \mathrm{m}.
$
%
%

Vertical bending will produce synchrotron radiation,
which, in its turn, will add to the vertical emittance
of the beam.  At high energy, the total energy radiated by one electron
is given by the expression,\footnote{
For example, see Equations 8.6 and 8.10 
of Edwards and Syphers. \cite{edw:andsyp}
}
\begin{displaymath}
U = \int ( c dt ) \frac{1}{6 \pi \epsilon_o} 
        \left( \, \frac{e}{\rho} \, \right)^2
        \gamma^4
~~,
\end{displaymath}
\noindent
where, $\rho$ is the bend radius,
$\gamma$ is the relativistic \textmath{1/\sqrt{1 - (v/c)^2},}
and the other variables need no introduction.
In terms of the electron energy, $E$, and the bend angle, $\theta$, produced
by the quad over its length, $\ell$, 
\begin{equation}
U = ( 1.41 \times 10^{-5} \, \mathrm{m} \, \mathrm{GeV}^{-3}) 
              \cdot E^4 \theta^2 / \ell
\label{eq:j3rgb9}
~~.
\end{equation}
%
%
%
%
\par
Using the same parameters as before,
\textmath{\theta \approx 5.6 \, \mu \mathrm{rad};}
at the high energy end of the linac,
\textmath{E \approx 500}~GeV,
and \textmath{\ell \approx 1}~m;
our estimate of the total radiated energy 
(per electron per bend)
is about 28~keV.
Put another way, 
the ratio, \textmath{U/E \approx 6 \times 10^{-8}.}
%
%
It is inconceivable that such a small fractional change
in beam energy could seriously damage the emittance,
but we will estimate its effect anyway.
The additional invariant emittance 
due to synchrotron radiation is approximated as
%
%
%
\begin{equation}
\Delta ( \gamma \epsilon_y / \pi ) 
\ \approx \ \frac{55}{6\sqrt{3}} \; \frac{r_e\hbar c}{mc^2} \left\langle
{\cal H} \right\rangle \left\langle \frac{1}{\rho^2} \right\rangle
 \; \theta \gamma^6 \; ,
\label{eq:1ap0nv} \\
\end{equation}
where $\epsilon_y = \pi\sigma_y^2/\beta_y$ and $r_e$ is the classical electron
radius.  The quantity $\langle{\cal H}\rangle \approx D_y^2/\beta_y$ at the
quadrupole location.

%
Plugging in the same numbers as before, 
estimating \textmath{\beta_y \approx 60}~m,
and using our previous estimates
for $U$ and $D_{\rm max}$,
we obtain,
\textmath{\Delta ( \gamma \epsilon_y / \pi ) \approx 1.8 \times 10^{-7}}~nm
-- as expected, a very small number.
%
\section{Localized sharp bends}
Although the 2~$\mu$m offset of an off-momentum particle predicted in the
previous section is not catastrophic, neither is it completely negligible.
We will now consider eliminating it by employing the second scenario:
constructing a Main Linac that is laser straight
except for highly localized bends at a few, widely separated
locations.
These bend sites then allow us to follow the equipotential
in a coarser, piece-wise fashion.
If we think of bending every kilometer, or so, then
the bend angle should be about 160~$\mu$rad. 
We'll take this as the ``canonical'' value for 
calculations in this section.
\par
We will proceed again in a minimalist way.  In order to minimize the
modification of existing lattice hardware and optics, we adopt the use of
combined function magnets to both bend and focus the electron beam.  Other
possibilities could be considered at a later date.
%
%
If we only ``bend'' at each of the local sites, a dispersion wave
of amplitude $\sim$~4~mm would be generated at each bend center.  To match
the trajectories at the end of each local bending region for particles with
various momenta, the total bend angle of each region is distributed  across
four neighboring (combined function) dipoles.  The strategy is akin to that
of an 18$^{\rm th}$ century optician designing a simple focussing achromat. 
\par
\begin{figure}[ht]
\centering
\includegraphics*[width=90mm]{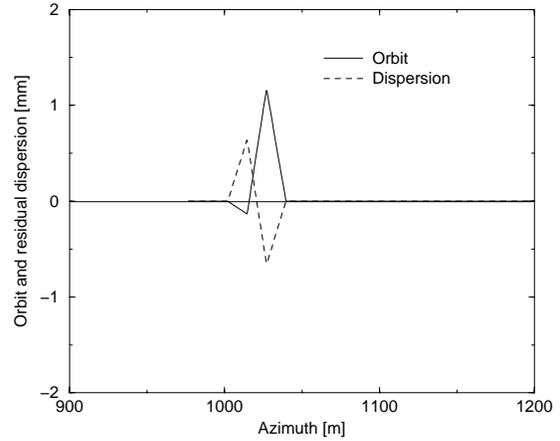}
\caption{
\label{fig:1b}
Orbit deviation required to zero the residual dispersion.
        }
\end{figure}

The results at the lowest energy bending location are shown in \fig{1b}.
The dashed line follows the residual dispersion,
now completely contained within the 
$\approx$~40~m long bending region,
with maximum amplitude of about 0.6~mm. 
The maximum orbit distortion of 1~mm is too large
an offset from the central (curved) axis of 
the local bending magnets.
They would have to be displaced so as to follow the new orbit.
A few iterations of these manipulations 
should then converge on an acceptable design.
However, the final orbit and its local residual dispersion
should not be much different from what we have calculated here.
For now, we simply display these results as indicating
the order of magnitude of the effects.
\par
\begin{figure}[b]
\centering
\includegraphics*[width=90mm]{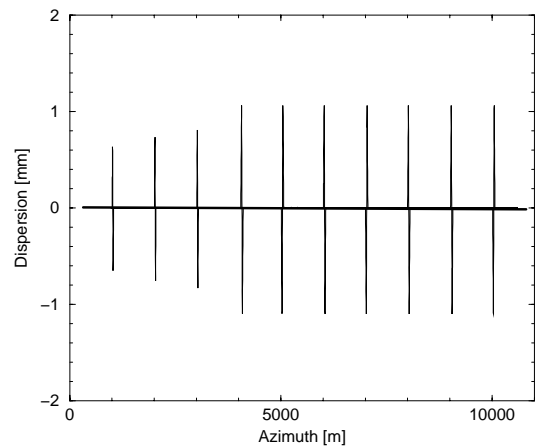}
\caption{
\label{fig:gib7vnb}
(a) Residual dispersion generated by the partial
achromats at all ten locations along the NLC.
        }
\end{figure}

The distortions in the trajectory at ten locations 
located $\sim$1~km apart in the NLC are of the order 1-2~mm, with the higher
displacements occuring at the higher energies.   The corresponding dispersions
generated along the linac are shown in \fig{gib7vnb}.
A residual dispersion of 1~mm remains in the 
neighborhood of the bends.
Again assuming that \textmath{\Delta p / p \approx 0.02,}
we have
$
D_{\rm max} \cdot \frac{\Delta p}{p} = 20 \, \mu \mathrm{m}
$
compared to
$\sigma_y = 
\sqrt{\beta_{\rm max} \epsilon_y / \pi} = 4.6 \, \mu \mathrm{m}.
$
\noindent
This is a large increase, but \emph{it exists only
near the bend sites.}
Away from these sites, the dispersion is (essentially) zero,
and its contribution to emittance is negligible.
We note in passing that an advantage of this calculation
is that one can envision making it operational.
%
%

\begin{figure}[h]
\centering
\includegraphics*[width=90mm]{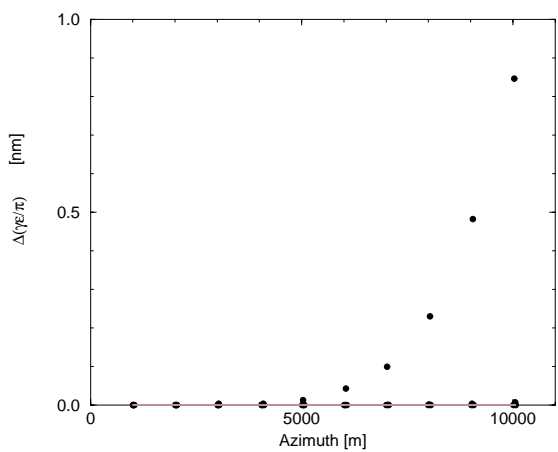}
\caption{
\label{fig:wx}
Emittance growth due to synchrotron radiation in sharp bends.
        }
\end{figure}

Finally, we estimate the synchrotron radiation
and emittance growth incurred by our second scenario,
once again using 
\eqs{j3rgb9} and (\ref{eq:1ap0nv}).
The values of \textmath{\Delta ( \gamma \epsilon_y / \pi )} 
at all bend locations are plotted in \fig{wx}.
Each site contains one dominant, very sharp bend.
Its effect is most apparent near the high energy end of the linac,
where the $E^6$ dependence becomes overwhelming.
Even so, the additional $\approx$~1~nm in invariant emittance 
is less than 1\%
of the 140~nm vertical emittance expected within the interaction region.
%
\par
Notice that although the synchrotron radiation is rather high
at the end of the linac, the ratio
$
U / E = 49 \ \mathrm{MeV} / 473 \ \mathrm{GeV} \approx 10^{-4}
~~
$
\noindent
is still a small number.
\par
In passing, we note that some attention should be given to coherent
synchrotron radiation (CSR) from individual bunches.  A quick
look\cite{mich:FNAL690} shows that the current NLC design has an aperture 
\mbox{$a\approx 7$ mm,} so CSR is forbidden, even at the high energy end of
the linac, because of ``shielding'' from the walls of the beam pipe.
However, the margin of safety is not comfortably large. It may be necessary
to reexamine this issue.
\section{Ultimate energy of \\ ``curved'' linac}
While present NLC designs with beam energies in the range of a few hundred
GeV to 1 TeV may not be very sensitive to the curvature of the earth, there
will be a practical limit to the upgraded energy of such a device.  Suppose
the scheme of steering the beam with offset quadrupoles is adopted.  Then,
at high energies, eventually the energy gain within a FODO cell will be
equal to the energy lost due to synchrotron radiation as the beam is bent by
the offset quadrupole.  This limit can be easily written as
\[  E_{lim} = \left( \frac{\pi}{C_{\gamma}} \; \frac{\ell_q}{L} \;
   E_{cv} R^2 \right)^{1/4}  \]
where $E_{cv}$ is the energy gain per meter of the linac.  As the energy is
increased, both the quadrupole length (strength) and the half-cell length
increase roughly proportionally.  As an example, using $E_{cv}$ = 50~MeV/m,
and $\ell_q/L$ = 0.025, then we get
$E_{lim} \approx $ 6.5~TeV.

\begin{figure}[t]
\centering
\includegraphics*[width=85mm]{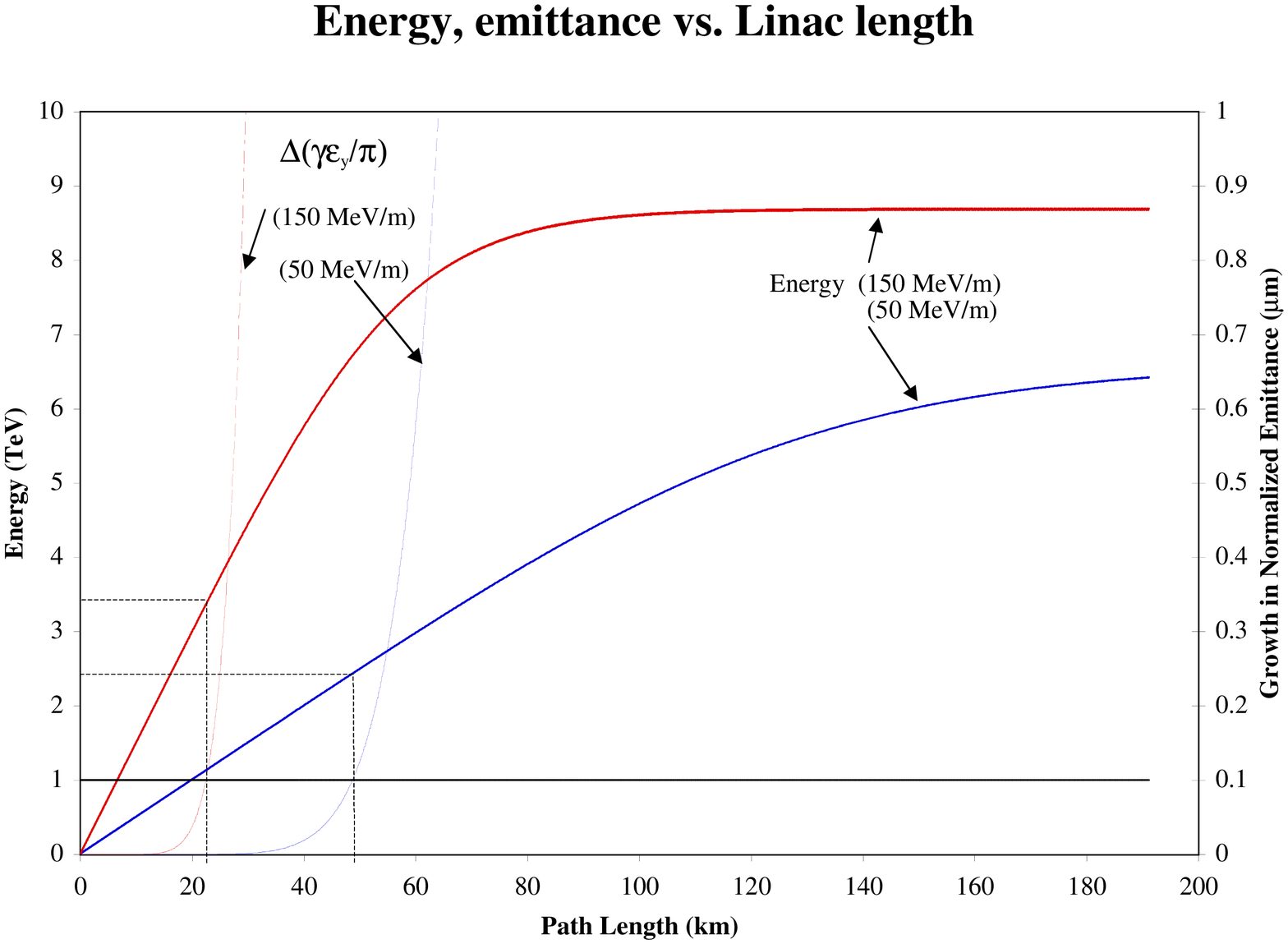}
\caption{
\label{f:CurvedLC}
Beam energy and emittance growth vs. path length of linac which follows the
curvature of the earth.  Results with accelerating gradients of 50~MeV/m and
150~MeV/m are shown.
        }
\end{figure}

This conclusion is illustrated in Figure~\ref{f:CurvedLC}.
The emittance growth due to synchrotron radiation is also plotted, showing
that a growth of $\Delta(\gamma\epsilon_y/\pi)$ = 100 nm (roughly equal to
the NLC nominal design value at collision) occurs well before the final
energy is reached, with the emittance growth being a steep function of energy
($\sim
\gamma^6$).  For 50~MeV/m, the practical limit of a curved linac may be only
about 2-3~TeV per beam, which would have a length of about 50 km.  For
150~MeV/m, the limits are 3-4~TeV per beam over about 20~km.  Note that a
``laser straight'' linear collider with 20~km per linac, and a lengthy
interaction region, with its two ends near the surface of the earth would
have its collision point located roughly 50~m below the surface.

%
%
\begin{center}
{\bf Acknowledgements}
\end{center}
We are grateful to Courtlandt Bohn for suggesting the possible importance of 
coherent synchrotron radiation.
\setlength{\itemsep}{0in}
\bibliographystyle{plain}

\end{document}